# Individual Bibliometric Assessment @ University of Vienna: From Numbers to Multidimensional Profiles

Juan Gorraiz, Martin Wieland and Christian Gumpenberger

juan.gorraiz, martin.wieland, christian.gumpenberger@univie.ac.at
University of Vienna, Vienna University Library, Bibliometrics and Publication Strategies,
Boltzmanngasse 5, A-1090 Vienna (Austria)

**Abstract**

This paper shows how bibliometric assessment can be implemented at individual level. This has been successfully done at the University of Vienna carried out by the Bibliometrics and Publication Strategies Department of the Vienna University Library. According to the department's philosophy, bibliometrics is not only a helpful evaluation instrument in order to complement the peer review system. It is also meant as a compass for researchers in the 'publish or perish' dilemma in order to increase general visibility and to optimize publication strategies.
The individual assessment comprises of an interview with the researcher under evaluation, the elaboration of a bibliometric report of the researcher's publication output, the discussion and validation of the obtained results with the researcher under evaluation as well as further optional analyses. The produced bibliometric reports are provided to the researchers themselves and inform them about the quantitative aspects of their research output. They also serve as a basis for further discussion concerning their publication strategies. These reports are eventually intended for informed peer review practices, and are therefore forwarded to the quality assurance and the rector's office and finally sent to the peers.
The most important feature of the generated bibliometric report is its multidimensional and individual character. It relies on a variety of basic indicators and further control parameters in order to foster comprehensibility. Researchers, administrative staff and peers alike have confirmed the usefulness of this bibliometric approach. An increasing demand is noticeable. In total, 33 bibliometric reports have been delivered so far. Moreover, similar reports have also been produced for the bibliometric assessment of two faculties with great success.

# 1. Introduction and purpose

In this paper we present an approach how bibliometric assessment has been implemented at individual level at the University of Vienna. This model has already been recognized and discussed at several occasions in different countries, and due to an increasing demand it is herewith made available on popular demand.

Bibliometric assessment is generally the responsibility of the Bibliometrics and Publication Strategies Department of the Vienna University Library (in the following referred to as Bibliometrics Department). Since its launch in 2008, the mentioned department has already successfully completed 33 individual reports (Gumpenberger et al., 2012).

It is important to emphasize, that the tasks of the department are not only restricted to support university administration in their research assessment exercises, but also include supportive services for the scientists themselves. Choosing the most successful publication channels is particularly important for the careers of young scientists.

Our Bibliometrics Department is committed to provide tailored services for these two beforehand mentioned target groups, the academic administration (rector's office and quality assurance department of our university, and the scientists themselves. Our primary concern is the prevention of "quick and dirty" bibliometrics and its consecutive incorrect and even harmful interpretations. We rather aim to achieve a situation with a well-informed administration on the one hand, and well-prepared scientists who can successfully cope with all these evaluation practices on the other hand.

The described approach was initially designed as a supportive bibliometric report for individual scientists. Such reports are intended to inform the scientists about the quantitative aspects of their research output and to serve as a basis for further discussion concerning publication strategies. Based on this service the scope has been expanded to an individual assessment of professors. At the University of Vienna, some professors need to undergo evaluation five years after their appointment due to the terms of their contract. Depending on the discipline the rectorate decides whether or not the Bibliometrics Department should provide its bibliometric expertise. Such reports are only complementary to the professors' self-assessments and are always generated and finalised in mutual agreement between the Bibliometrics Department and the professors to be evaluated. Once finalised these reports are checked by the Quality Assurance Department, then forwarded to the rector's office and finally sent to the peers. By this means the latter hopefully refrain from performing inadequate bibliometric analyses and rather focus on the qualitative assessment (Weingart, 2005; Bach, 2011; Glänzel & Wouters, 2013).

The most important feature of a bibliometric report is its multidimensional and individual character. For each individual a personally elaborated report is tailored according to corresponding research field(s). This process includes the selection and use of the adequate data sources, the consideration of different publication cultures and publication channels, and the appropriate use of the available tools for analysis and presentation.
We are convinced that individual evaluation requires personalized treatment and cannot be achieved by automatized "push the button" evaluation reports. However, many of the currently available analytical tools are indeed helpful to optimize and accelerate proper individual assessment.

Indeed, bibliometric analyses should never rely on only one particular indicator, since this normally means a restriction to only one aspect. In spite of the fact that composite indicators aim to combine several aspects, they rather complicate than simplify the interpretation of the results for the addressed target group. Therefore, our approach relies on a variety of basic indicators and further control parameters, in order to do justice to the multidimensionality of the problem and to foster comprehensibility.

**2. General structure of the individual assessment**

The individual assessment comprises of the following steps:

1. Interview with researcher under evaluation
2. Report or "bibliometric profile" of the researcher's publication output
3. Discussion and validation of the results with researcher under evaluation
4. Optional analyses

Each of these steps is described and discussed in the forthcoming sections.

**2.1. Interview with researcher under evaluation**

This is one of the most important and most relevant parts of the individual assessment. On the one hand, the bibliometricians and evaluators gain valuable insight into the researcher's work and the peculiarities of the corresponding research field. On the other hand, the researcher gets an opportunity to understand the applied evaluation methods and tools and to discuss suitability and restrictions.

Scientists tend to be busy and certainly cannot spare too much time for interviews. In order to stress the importance and to guarantee availability, the rector's office invites the researchers to participate actively in the evaluation process.
Interviews can last from one to two hours (at most) and generally take place at the researchers' workplaces. Following questions are always asked:

1. Which data sources do you use regularly for retrieving literature in your research field? Do you use alert services?
2. Do you use permanent person identifiers (like ORCID, ResearcherID, etc.)?
3. Do you have a complete record in our CRIS - Current Research Information System?
4. Do you use repositories?
5. Have you submitted preprints in order to claim priority?
6. What are the most important publication channels in your field (special emphasis on monographs, book chapters, patents if appropriate)?
7. Does the order of authors (first, last or corresponding author) play a role in your research field? If no, why not?
8. Which criteria are relevant for your publication strategy?
9. Is Open Access also a valid criterion according to the recommendation of our university? If no, why not?
10. Do you actively participate in conferences?
11. Are you an editor of one or more scientific journals? If yes, which ones?
12. Do you actively support the peer-review system by providing reviews? If yes, how many per month?

13. Do you maintain a personal website? An entry in Wikipedia? A Google Scholar Citations profile?
14. Do you use a reference manager system? If yes, which one? Why do you think it is helpful?
15. Do you actively engage in mailing lists or blogs? If yes, which ones?
16. Do you use other Social Media tools? If yes, which ones?
17. What do you think about usage metrics (downloads) and altmetrics?
18. Do you generate research data? If yes, how do you manage and archive them?
19. How do you (or would you) select and assess colleagues or potential collaborators? If yes, do you also embrace quantitative methods?
20. Is there anything else we have not covered so far and you would like to share?

Questions 1 and 6 are crucial for the selection of the data sources used for the bibliometric analyses, whereas questions 2 and 3 are relevant for data disambiguation. Question 7 informs about the need for such an analysis. However, this will also be checked in the databases independent from the interviewee's feedback.
Questions 4, 5, 8 and 9 are relevant for the design of the visibility analysis.
Questions 10, 11 and 12 inform about the researcher's experience and reputation in the field.
Questions 13 until 17 are relevant in order to learn about the researcher's attitude to new metrics and social media.
Question 18 has been included since research data management is an emergent topic (Costas et al., 2012; Bauer et al., 2015).
Responses to questions 19 and 20 finally allow to meet the researcher's particular expectations.

**2.2. Bibliometric report**

Publication data are provided by the researcher under evaluation in form of a publication list as agreed in the previous interview. The list is compared to the data retrieved in the bibliometric data sources by the bibliometricians and amended if necessary.

The resulting bibliometric report itself is custom-tailored for each professor according to individually relevant aspects and to the accepted publication culture in the according discipline.

The structure of the bibliometric report generally comprises of the following sections:

- Methodology
- Coverage in databases
- Activity analysis for publications
- Affiliation and funding analyses
- Co-authorship analysis
- Visibility analysis
- Impact analysis
- Citing analysis
- Network and cooperation analyses
- Reference analysis
- Research focus
- Summary
- Annex

Each section will be described in full detail to foster a better understanding of our approach. In order to round off the information an anonymized report is provided in the annex.

### 2.2.1 Methodology

The methodology section includes a thorough description of the databases and indicators selected for the bibliometric analysis.
As it is already a well-established practice at the University of Vienna (Gumpenberger et al. 2012: Gorraiz et al. 2015), the bibliometric standard analyses are meant to shine a light on three different main aspects:

- Activity: the number of publications along a timeline and with differentiation of document types to reflect the productivity (Lotka, 1926). Furthermore, authorship and affiliation analyses (like number of co-authors or author's role) are provided as well (; Shockley, 1957).
- Visibility: the percentage of publications indexed in well-respected databases (see coverage) as well as the prestige and impact of the journals where the researcher has published in, according to the Journal Impact Factor (Garfield 2005; Glänzel and Moed 2002) or other alternative journal impact measures such as SCImago Journal Rank (SJR) (Gonzalez-Pereira et al. 2009) and Source Normalised Impact per Paper (SNIP) (Moed; 2010, 2011), in order to reflect the editorial barrier and to unveil publication strategies.
  Visibility plays a key role whenever the evaluation covers only the most recent years. In this case, the citation window is too short and the relevance of such citation analyses is limited. Furthermore, higher visibility increases the chance to be cited.
- Impact: a citation analysis including several indicators to reflect the significance in the scientific community (Cronin 1984; Van Raan, 2004; Moed 2005; de Bellis 2009; Vinkler 2010).

Tables 1 and 2 inform about the different aspects of a bibliometric profile and its corresponding indicators.

Additionally, an analysis of the citing documents (see impact table 1), of the research focus and interdisciplinarity (see Focus), of the cooperation networks at different levels (see Table 2) and of the cited references (see Knowledge Base) are provided. "Other metrics" and "Self-marketing in Internet" are discussed in the sections "Interview" and "Optional Analyses".

It cannot be stressed often enough that citations are only used as a proxy for the impact (and not for the quality) of the publications in the "publish or perish" community (i.e. the researchers who are committed to publishing their results).

Visualization is done with the freely available software packages BibExcel (Persson et al, 2009), Pajek (De Nooy et al., 2005) and VOSviewer (Van Eck & Waltman, 2010). In the resulting maps the size of the circles is proportional to the number of publications, whereas the width of the lines is proportional to the strength of their co-occurrence.

**Table 1: Dimensions and indicators**

| Activity | Visibility (publication strategies) | Impact (Citations) | Focus |
|---|---|---|---|
| # Publications & Trend lines | # indexed in Databases (coverage) | # citations (total, mean, maximum) | maps based on titles & abstracts, descriptors, keywords and identifiers |
| # Document Types | # & % English | Normalised Citation Score (CNCI, Crown-Indicator) | interdisciplinarity according to Subject Categories |
| # Authors (Mean, Maximum, # single-authored) | # in Tops Journals (according IF. SJR, SNIP or journal's lists or rankings) | # & % Tops in Percentiles (Top 1%, Top 10%) | |
| # Author's role (first, last, corresponding) | aggregate & median category impact factor | h-index & variations (g,m); i-indices | |
| # patents | # Open Access | % self-citations | |
| # research data sets ? | books ? | analysis of citing documents | |

**Table 2: Dimensions and indicators (Part 2)**

| Cooperation | Other Metrics | Knowledge Base | Self-marketing in Internet |
|---|---|---|---|
| based on Affiliations: intensity (# publs) & impact (# cits, cits/publ, CNCI, % Top10%, %Top 1%) | usage metrics: views & downloads | reference analyses (cited documents) | in repositories |
| % international collaboration % domestic collaboration % industry collaboration | altmetrics (captures, mentions, social media, etc) | state-of-the-art (PY of cited documents), most cited document types, most cited journals | in Google Scholar, in Wikipedia |
| network analyses at different levels (scientists, institutions and countries) | | benchmarking with other leading scientists in the same research field | in mailing lists, blogs, reference managers and other social media |

## 2.2.2. Coverage analyses

Coverage analyses have two main purposes: first, to select the adequate data sources for the forthcoming analyses, and second, to shed light on the visibility of the research performance.
This second aspect is based on the fact that publications indexed in international renowned databases are more visible than the non-indexed ones, and that they can be retrieved more easily.

The Web of Science (WoS) – Core Collection is used as the preferred data source for bibliometric analyses, since being indexed in this database is generally perceived as a sort of "high impact" (or at least high visibility) criterion within the scientific community. All the faculties related to the natural sciences have corroborated this perception.
Analyses are performed in the source part as well as by using the cited reference search, especially for other document types than contributions in journals.
Due to the fact that not all disciplines are equally well covered in WoS, alternative data sources such as Scopus or Google Scholar are used for complementary analyses.

Scopus is used as second citation database, in order to avoid or correct indexing errors in WoS and to benefit from the larger number of indexed journals (almost twice as many as in WoS).
Google Scholar (GS)[1] via "Publish or Perish"[2] (Harzing, 2007) is considered as a complementary bibliometric source. It stands out because of its higher coverage for some publication types (like monographs, reports, etc.), which are more relevant in the social sciences and the humanities.

This set of data sources is always complemented with at least one subject specific database. The choice is made based on the preference of the researcher under evaluation. Such popular additional databases are, for instance, Chemical Abstracts or Mathematical Reviews.

However, for citation analyses only products are considered that include the corresponding metrics or at least citation counts.

## 2.2.3. Activity analyses for publications

The first activity analysis is performed according to the publication list provided by the researcher under evaluation itself.
The most important publication types and document types are identified.

Document types used by the authors in their publication lists are manually reassigned to these generally included standard groups: monographs (books), book chapters, journal articles, proceedings papers, conferences (including meeting abstracts and talks), book reviews, edited books and journal issues, and other publications (or miscellaneous). Reports or working papers and patents are included whenever appropriate, mostly for disciplines related to physics, the life sciences or technology.
Some publication types occasionally receive special attention according to their disciplinary importance, such as proceedings papers in computer sciences or book reviews in the social sciences.

---

[1] Analyses in GS should be taken with a pinch of salt. GS is rather a search engine than a database, and therefore indexing remains non-transparent and documentation is lacking.
[2] 'Publish or Perish' is a software programme that retrieves and analyses academic citations. It uses GS to obtain the raw citations (see also, http://www.harzing.com/pop.htm)

Different document types and publication windows are distinguished in the results of the activity analysis. The standard analysis contains a chart providing the evolution of the past ten complete years. This is done for all document types as well as for the most important document types (articles, citable items, etc.). Information about earlier years or the most recent uncompleted year is provided separately.

The activity or productivity is measured by means of absolute output values – that is, normal counts. In order to relativize the obtained results, complementary co-authorship analyses are performed.

Data automation is desirable, but currently no automation can deliver the same reproducible results. Therefore we attach special importance to the degree of coverage in the databases used for our analyses and match the obtained search results with the provided publication lists whenever possible. Automation will gain momentum once a critical mass of permanent individual identifiers (like ORCID) has been implemented within the scientific community.

### 2.2.4. Co-authorship analysis

The total number, the average number (mean and median) and the maximum number of co-authors are determined for different periods in order to analyze their progress in time (Laudel, 2002; Glänzel, 2014). Furthermore, the number and percentage of single authored publications as well as the author's publication role (number and percentage of publications where the researcher is first, last and/or corresponding author) are studied for different periods.

The order of authors is mostly determined by the degree of contribution, but can also by alphabetical in some fields. The initial interview with the researcher under evaluation sheds light on this issue, and all provided information is easily corroborated by the bibliometric analysis.

"Co-author dependence" (i.e. percentage of publications with the same co-author) is always reported especially when it exceeds 75%.

### 2.2.5. Affiliation and funding analyses

Correct affiliation information enhances institutional visibility and directly influences the position in university rankings. Most rankings rely on data from WoS or Scopus. Therefore, affiliation analyses are usually performed in these databases.
Affiliation changes and how these might affect the productivity of a researcher are also considered in this type of analysis.

Funding analyses are also performed in order to inform about the number and percentage of funded publications as well as the main funding agencies. These analyses are performed in WoS and Scopus and offer quite reliable results since 2008.

### 2.2.6. Visibility analysis

The visibility analysis comprises of two parts: first, the number and percentage of publications indexed in the different international, well-respected selected data sources as already mentioned under coverage analysis, and second, the number and percentage of publications in top journals or sources.

The visibility of a document is determined by the reputation or the impact of the source where it was published. It reflects the editorial barrier and unveils publication strategies.
Therefore, the journals or sources where the researcher under evaluation has published in are analysed and compared for different time periods.
For a journal article, the visibility can be determined by the impact measures of the journal it was published in. The most common impact measure is the journal impact factor (IF) (Garfield, 2005; Glänzel and Moed, 2002). Thus, a document has a high visibility in one research field, if it was published in a journal with an IF bigger than the aggregate or the median IF of the corresponding subject category or field. Therefore, visibility can be quantified by the IF of the source in relation to the aggregated or median IF assigned to the corresponding subject category.
The IF is an appropriate visibility measure, but only for journals indexed in the Journal Citation Reports (JCR). Other recent alternatives are based on the widely known PageRank algorithm of the Google search engine – for example, the article influence score or the SJR indicator (Gonzalez-Pereira et al., 2009). SJR and SNIP (Moed, 2010 & 2011) refer to journals indexed in Scopus, which results in a considerable increase of "visible" journals to almost 21,000 journals.

The JIF of the most recent JCR edition is used for all analysed publications as an accepted compromise. [3] In our approach, we predominantly use IF quartiles. The quartiles (Q1 = Top 25%, Q2 = Top 25-50%, Q3 = Top 50-75%, Q4 = Top 75-100%) in the corresponding Web of Science category[4] are calculated based on the IF data reported in the last available edition of the Journal Citation Reports (JCR) in the corresponding Web of Science category (impact of the journal at the time of the evaluation).
Due to the fluctuations of the IF, discrepancies are expected according to the method employed. However, the use of quartiles addresses these shortcomings significantly, because the quartiles are less volatile (Gorraiz et al. 2012b).

The visibility analysis includes the list of all journals and serials, where the scientist under evaluation publishes in. Following aspects are taken into account:

1) The number of items published in the last ten years
2) The number of citations attracted by the publications in each journal
3) The corresponding journal impact measure (IF, Article Influence Score, SNIP, SJR)
4) The corresponding quartile according to the selected journal impact measure.

Furthermore, the allocation of all publications (2005-2015) to the different IF quartiles and the comparative quartiles distribution for the publications either published in the interval 2005-2009 or 2010-2014 are calculated and plotted in figures. These analyses are performed in order to reveal considerable changes in the publication strategy and journal preferences in the previous five years.

---

[3] Another possibility would be to consider the IF in the JCR edition corresponding to the publication year. However, this method is also not completely correct, since the JCR edition is always one year delayed and the calculation of the IF considers either the two years or five years prior to the publication year (Gorraiz et al., 2012b). A third possibility would be to use the mean value of the impact factor from all publication years.
[4] If the journal has been assigned to several WoS categories, the best quartile is used. This decision aims to help the researcher, who could always argue that multiple assignments are discriminatory.

In disciplines where the coverage in WoS and Scopus is known to be low, such as in the social sciences, mathematics and the computer sciences, committees and faculties have the possibility to provide self-compiled lists of "highly" reputed journals for their discipline. In such cases, the number of publications in these selected journals is calculated.

Unfortunately, it is much more difficult to assess the visibility of monographs (see Table 1). Such analyses for publication types like edited books or monographs are highly controversial. Reputation of the editorial board, circulation, number of editions, holdings and loans in international catalogues are the most relevant indicators suggested, but none of these so-far suggested approaches has proven to be suitable for research assessment purposes.

**2.2.7. Impact analysis**

Impact finally relies on citations as proof of recognition within the scientific community. Citation analyses for publications in journals are commonly performed in the source part of WoS. However, the "cited reference search" is also used in order to collect citations to other document types that are not indexed in the source part of WoS, particularly if these document types are common publication channels of the researcher under evaluation.

In order to consider the skewness of most of the citation distributions (Seglen, 1992), three indicators are used to describe the distribution of citations: the total sum, the arithmetic mean and the maximum[5]. Furthermore the number or percentage of cited documents is considered and the arithmetic mean substituted by the number of citations per cited document, which is a more significant indicator. Moreover, the h-index is determined for all document types (Hirsch, 2005; Bar-Ilan, 2008; Alonso et al., 2009).
Citation analyses are performed for citable items (article, reviews and proceedings papers) as well as for all items (Gorraiz and Gumpenberger, 2015). The percentage of self-citations (Glänzel et al. 2004 & 2006) is calculated and included as "Control Data". Values below 20% are considered as usual, whereas higher ones have to been explained.

Citation counts are an accepted proxy for impact. However, normalisation is needed according to discipline and per publication year (Schubert & Braun, 1986 & 1996; Costas el al., 2009). Our multifaceted approach is based on the usual indicators (citations, citations per cited publication, maximum of citations, h-index and g-index), but also incorporates normalised citation counts in the form of the "Category Normalized Citation Impact" (CNCI)[6] and the number and percentage of Top 10% and Top1% most cited publications (Adams et al. 2007; Gorraiz et al. 2011, 2012a; Bornmann et al. 2012). Top 10% is used in order to assess the degree of excellence and Top 1% allows a further differentiation between highly cited ("excellent") and extremely highly cited publications ("edgy" publications).

CNCI, Top 10% and Top1% most cited publications can be calculated according to the ESI percentiles (22 categories) or by using the tool InCites, which enables different calculations to include other classifications or the corresponding WoS categories (more than 250). In this latter case, fractional count is used when the journal is assigned simultaneously to different WoS categories.

---

[5] The standard deviation is provided only upon request.
[6] The "Category Normalized Citation Impact" (CNCI) provides the citation impact (citations per paper) normalized for subject, year and document type and is calculated according to the data collected via InCites. It is also named "Crown Indicator" or "Field Citation Score".

Citation analyses are mostly performed in WoS Core Collection, Scopus and in at least another subject specific database with included citation counts (like ADS, HEP, Mathematical Reviews, Chemical Abstracts, BIOSIS, etc). Google Scholar (via "Publish or Perish" and/or Google Scholar Citations Profiles) has so far been used in an exploratory way for the humanities and the social sciences.

Field normalized indicators based on reference values are not available in WoS Cited Reference Search or in Google Scholar (data sources not providing reference citations values per subject category and year of publication) and were then substituted by different variations of the i-index (some variations of the i-index, starting with the i10[7], i50, i100, and i100), according to the number of citations attracted in each source. It should also be considered that the "i-index" thresholds are determined according to the expected number of citations for each discipline. Therefore, in the social sciences, only the i10-index and the i50-index are common.

Citation analyses for monographs relying on both the book citation index and the "cited reference search" in WoS are performed separately in order to avoid inconsistencies by mixing different metrics (Gorraiz et al. 2013).
If appropriate and desired by the researcher under evaluation and in consideration of the previous interview, further document types such as patents, e-publications, articles in newspapers, etc. are also taken into account. Patent analyses are performed in Espacenet at the EPO, or for some fields (like chemical, engineering, electrical and electronic and mechanical engineering) in Derwent Innovations Index or in CAS (Chemical Abstracts).

### 2.2.8. Analyses of the citing documents

All citing articles are retrieved in WoS using the citation report. Another possibility is to enlarge the analysis to the Cited Reference Search in order to include also citations to non-indexed publications in Web of Science Core Collection.

Mainly two analyses are performed. First, the citing countries are determined. A network map of the citing countries informs about the degree of internationalisation concerning the impact of the researcher under evaluation (e.g. see Figure 1).

The size of the circles is proportional to the number of publications; the width of the lines is proportional to the strength of their co-occurrence.

Second, the citing publications can also be analysed according to:

a) Their visibility: percentage of top journals citing publications of the researcher in evaluation
b) Their impact: CNCI, percentage of Top 10% and Top 1% most cited among the citing documents.

---

[7] Number of publications with at least 10 citations

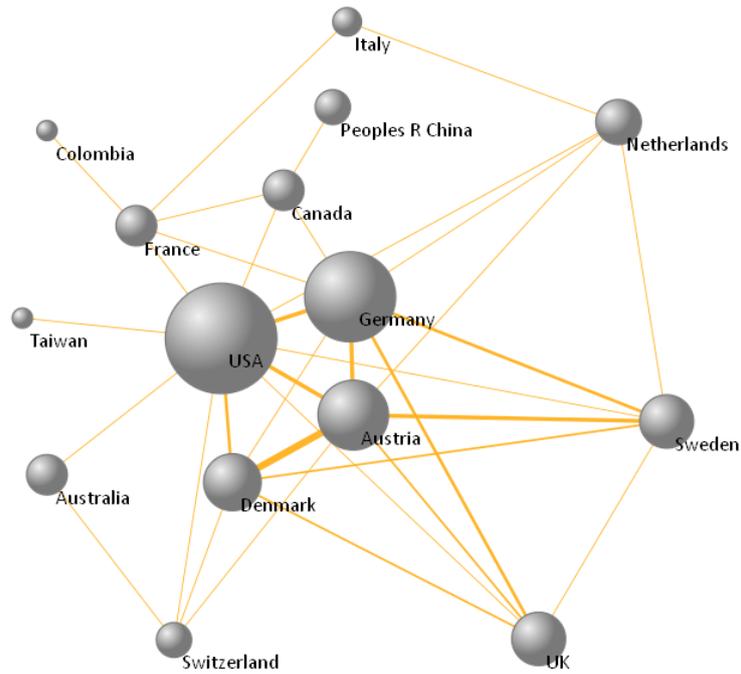

*Fig. 1: Country map of citing publications*

### 2.2.9. Cooperation analyses

Primarily, the proportions of international, national and domestic collaboration and their time evolution are analysed (Persson et al., 2004).
Further analyses are then performed at country, affiliation and author levels.

### 2.2.9.1. Cooperation on country level

An international network on country level is shown in this example of a corresponding network map (see Fig. 2).

This cooperation map is compared with the impact map created in section 2.2.8 and clearly shows that impact is normally much broader than pure collaboration. Thus the map representing the citing countries has more vertices and a higher density. In principle, this analysis could also be performed on affiliation or author level (considering e.g. citing institutions versus cooperating institutions, see also 2.2.9.2).

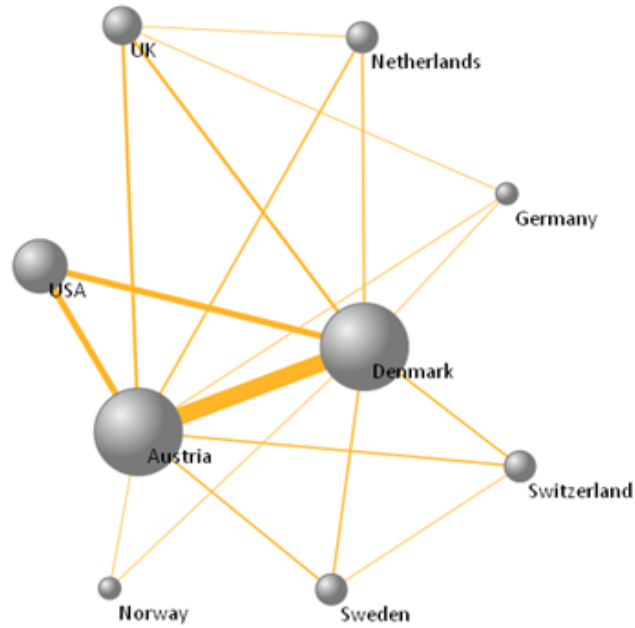

*Fig. 2: Co-publication network at country level*

### 2.2.9.2. Cooperation on affiliation level

An overview of the most cooperative institutions is provided in a table, which includes institution name, country name, the number and percentage of shared publications, the number of citations attracted, the CNCI (see 2.2.7), the percentage of Top 10% and Top1% most cited publications, the percentage of international collaboration and the percentage of collaboration with industry. Most of these data are obtained via InCites.

The number and percentage of shared publications informs about the volume or intensity of the cooperation, the number of citations attracted about the total impact of the collaboration, the CNCI and the percentage of Top 10% and Top 1% most cited articles about the mean impact and the excellence respectively.
The produced table shows that the most collaborative institutions (highest number of co-publications) are not always responsible for the highest CNCI scores and the highest percentage of top publications.

### 2.2.9.3. Cooperation on author level

A map informs about the network on author level. The number of different co-authors as well as the most collaborative authors can be identified.

### 2.2.10. Reference Analyses

The reference analyses inform about the knowledge base of the researcher under evaluation. They reveal which sources have been used and cited.
The total number of cited references, the percentage of cited journals or serials and the percentage of citations to other discipline-specific publication types are determined. These calculations are all performed by means of the software package Bibexcel as well as by further manual disambiguation.

Moreover, the state-of-the-art (publication years) of the cited references is represented in a figure. The publication years of the cited references are then compared with the cited half-life in the corresponding research field. Figure 3 shows an example.

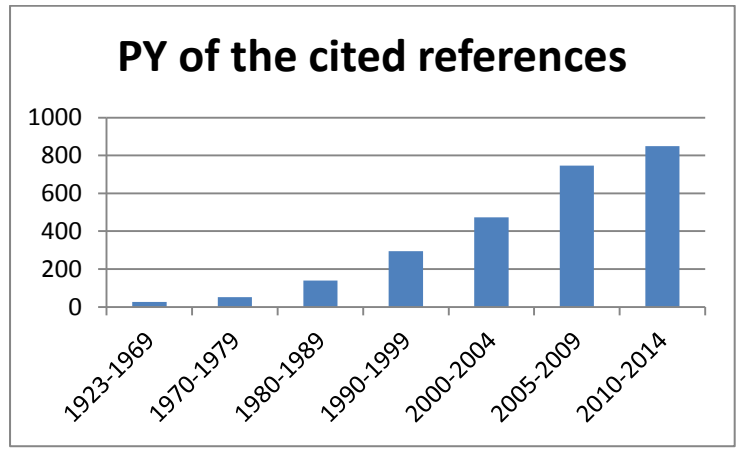

*Fig. 3: State-of-the-art (publication years) of the cited references*

The most cited sources and journals are determined, analysed (percentage of top journals) and compared with the journals, where the researcher under evaluation uses to publish in (see visibility analysis). A big match is a strong indication that the scientist under evaluation has managed to publish in the most relevant sources of his research area.

This analysis also helps to identify the most relevant sources in highly specific research fields, which is useful for correcting apparent limitations of subject classifications.

### 2.2.11. Research focus

Term-based co-occurrence maps are helpful to identify the most important research topics and the research focus of the researcher under evaluation. Terms are extracted from the title and abstract fields from WoS or Scopus publications. Co-occurrence maps can then be created with the tool VOSviewer (see Example, Figure 6.1). A minimum number of occurrences of a term is determined as a threshold. Subsequently a relevance score is calculated based on the field "times cited". According to this score, the 60% or 70% most relevant terms are selected.

More specific maps can be obtained by using controlled vocabulary (descriptors and identifiers) instead of terms extracted from the title and abstract fields. The resulting maps are compared and discussed.

### 2.2.12. Summary and Annex

The most important results are wrapped up in order to allow a fast overview. Moreover additional data are presented in an annex for transparency's sake.

### 3. Discussion and validation of the results with researcher under evaluation

The bibliometric report is always discussed with the researcher himself before it is handed over to the quality assurance, the rector's office and the peers. It complements the researcher's qualitative self-evaluation report and must not be taken out of this context.

## 4. Optional analyses

Last, but not least, three optional analyses are offered.

The first one is related to the research focus. With the researcher's help, his research field is delineated in one database, preferably in Web of Science or Scopus. The researcher is advised to save the resulting search string in order to create an alert service and to be automatically notified about any forthcoming relevant publications in his topic.
Based on the search results, the most important key actors of the last ten years are identified: most active authors and institutions, as well as funding agencies. Then the most cited publications and most cited first authors are retrieved and/or plotted in a map (bibliographical coupling). Finally, the top most cited publications (Top 10% and Top 1%) are identified.

The second analysis addresses the exploration of references. The researcher is asked to choose three leading peers in his research field (this can be based on the data retrieved from the previous optional analysis if available). The researcher's output will be then compared to the output of these selected peers for the last 10 years, including visibility, impact, cooperation and focus analyses as already pointed out before. Additional value is now created by doing a reference analysis. The most cited references are identified for the selected peers and then compared with the most cited ones of the researcher under evaluation. The researcher can then decide whether or not these sources are relevant for him, and if deviations are intentional or not. This type of analysis provides valuable insight on how to enhance one's knowledge base.

The third analysis is related to assessing the societal impact or the impact on the web (Galligan et al., 2013; Konkiel, 2013; Bornmann, 2014; Haustein et al. 2014). For this purpose, we currently explore two tools, altmetric.com and PlumX, sometimes even complemented by data retrieved from Scopus.
Publications are mainly analysed by using the DOI of each publication, but also by means of other options like the URL, ISBN, Patent number, etc. depending on the selected tool.
Most of our analyses were so far performed with PlumX. It allows a differentiation between citations (in Pubmed and Scopus), usage data (Ebsco, etc.), captures and mentions.

## 5. Retrospective overview of the bibliometric assessment

The usefulness of the bibliometric assessment has been confirmed by all the positive feedback obtained from researchers, research managers and peers on the one hand, and by an increasing demand on the other hand. After a pilot phase in 2010, the assessment service started in 2011 with eight bibliometric profiles and increased steadily since then to ten in 2014 and 15 in 2015. In total, 33 bibliometric reports have so far been delivered. For 2016, the rector's office has already commissioned more than 30 further reports.

Table 3 gives an overview of the research fields analysed. It shows that 75% of the bibliometric reports refer to research fields related to Natural or Life Sciences, and 25% to the ones related to Social Sciences.

Table 3. Overview of the research fields analysed

| Research Field | # Reports |
|---|---|
| Microbiology | 3 |
| Economics | 3 |
| Psychology | 2 |
| Computer Science | 2 |
| Astrophysics | 2 |
| Political Science | 1 |
| Mineralogy and Crystallography | 1 |
| Structural and Computational Biology | 1 |
| Botany and Biodiversity Research | 1 |
| Pharmacy | 1 |
| Environmental Geosciences | 1 |
| Sociology | 1 |
| Food Chemistry | 1 |
| Computational Physics | 1 |
| Gravitational Physics | 1 |
| Particle Physics | 1 |
| Inorganic Chemistry | 1 |
| Physics | 1 |
| Limmnology | 1 |
| Biophysical Chemistry | 1 |
| Zoology | 1 |
| Sport Science | 1 |
| Mathematics | 1 |
| Ecogenomics and Systems Biology | 1 |
| Meteorology and Geophysics | 1 |
| Material Physics | 1 |

The most used data sources were WoS Core Collection, Scopus and Google Scholar. According to the research field, other databases were also used, like: MathSciNet (Mathematics), ADS and Inspire-HEP (Astrophysics, Particle and Gravitational Physics), Sociological Abstracts and EBSCOhost (Sociology), ADS (Computer Science) and RePEc (Research Papers in Economics, including CitEc). As already mentioned above, the selection of the data sources was previously discussed and agreed with the scientist himself.

The bibliometric analyses were generally performed within two or three days by two colleagues from the Bibliometrics and Publication Strategies Department, depending on the number of considered data sources. Most of this time was spent on thorough data disambiguation and data cleaning. Candidates with implemented personal identifiers such as ORCID (Open Researcher and Contributor ID) or Thomson Reuters ResearcherID were definitely quicker to assess, provided that the profiles were regularly updated.

In the meantime, similar reports have been produced for the bibliometric assessment of two faculties with great success.

## 6. Lessons learned and conclusions

According to the philosophy of our department bibliometrics is not only a helpful evaluation instrument contributing to complement and reinforce the peer review system. It should also be perceived as a compass for researchers in the 'publish or perish' dilemma in order to increase general visibility and to optimize publication strategies.

This philosophy has proven to be valid throughout the whole assessment exercise.
The initial interviews as well as the follow-up discussions with the researchers under evaluation fostered a win-win situation for both, researcher and bibliometrician, alike.
All researchers could finally embrace the advantages of a thorough bibliometric report. Thus the bibliometric assessment exercise transformed for them from a nuisance to a valuable asset.
On the other hand we bibliometricians gained insight into many interesting aspects of a researcher's daily routine and learned much about discipline-specific publication habits.

Bibliometric expertise was not only appreciated by all researchers, but also by administrative staff and the peers. Its usefulness was generally confirmed and particularly emphasized for the life sciences. The positive experience gained in the social sciences is encouraging for our department to further explore suitable assessment procedures for researchers in the humanities.

Individual assessment is certainly complex and time-consuming. However, the valuable information produced can provoke positive changes at individual level, which in the long run are also beneficial for the institution itself. In terms of institutional visibility, the outcome can even be enhanced by policies (affiliation policy, publication strategy policy, open access policy, etc.), which – to come full circle - take effect at individual level.

Our suggested bibliometric report offers several advantages, which are highlighted below:

- It avoids complicated composite indicators, but rather relies on single indicators, which are particularly easy to understand for the researchers, the peers and administrative staff.
- Its multidimensional approach sheds light on various aspects, such as coverage, activity, visibility, impact, cooperation, research focus and knowledge base. It thus paints a diverse picture of a researcher's publication output.
- The practised inclusion and comparison of different data sources is helpful to identify and correct indexing and coverage errors.
- Finally, visualization (by means of network maps) helps to identify relevant clusters and fosters a better understanding of complex circumstances.

As a reputable service we do not only highlight the benefits but also the limitations of bibliometrics as an assessment method. It cannot be stressed enough that only quantitative aspects are measured in such a bibliometric assessment exercise. These are certainly objective per se, but should never be taken out of context. Each researcher is unique and has a particular history and individual skills. Just as each discipline has a particular publication culture. This should always be taken into account whenever peers set the course for the future career path of researchers. It is certainly irresponsible to exclusively rely on (unfortunately often practised "quick and dirty") bibliometrics and ignoring the big picture.

Last, but not least, our experience illustrates the crucial role modern scientific libraries are predestined to play in research assessment exercises. The field of bibliometrics is ideal for academic librarians to strengthen their on-campus position. Bibliometrics offers a wealth of opportunities to provide innovative services for both academic and administrative university staff. In so doing, librarians can actively contribute to the development of new publication strategies and the advancement of innovation.

**Acknowledgements**

The authors wish to acknowledge Ursula Ulrych, Steve Redeing and our Department for Quality Assurance as well as all the researchers under evaluation for their much valued support and input.